\documentstyle[aps,floats,graphicx]{revtex}

\begin{document}

\newcommand{\bec}{\begin{center}}
\newcommand{\ec}{\end{center}}
\newcommand{\be}{\begin{equation}}
\newcommand{\ee}{\end{equation}}
\newcommand{\beqn}{\begin{eqnarray}}
\newcommand{\eeqn}{\end{eqnarray}}
\newcommand{\bet}{\begin{table}}
\newcommand{\ent}{\end{table}}
\newcommand{\bib}{\bibitem}

\wideabs{

\title{
What is the real driving force of ion beam mixing?
}

\author{P. S\"ule$^1$, M. Menyh\'ard$^1$, K. Nordlund$^2$} 
  \address{$^1$ Research Institute for Technical Physics and Material Science,\\
Konkoly Thege u. 29-33, Budapest, Hungary, sule@mfa.kfki.hu/www.mfa.kfki.hu/~sule\\
$^2$ Accelerator Lab., Helsinki, Finland
}

\date{\today}

\maketitle

\begin{abstract}
Molecular dynamics simulations have been used to study the driving force of
ion irradiation induced interfacial mixing in metal bilayers in which the
relative mass of the constituents is considerable.
We find no apparent effect of chemical forces, such as heat of mixing
or cohesive energy up to 8 keV ion energy, although a considerable
number of liquid and high energy particles (hot atoms) persist up to even $20$ ps during the thermal spike.
This result is in direct conflict with the widely accepted theory of
thermal spike mixing (chemical interdiffusion model).
The supersaturation of vacancies also occurs and which induces a thermally activated
intermixing of the lighter constituent of the bilayer.
The delay and the decoupling of the intermixing of the light constituent is explained
as a backscattering effect at the interface:
the interface acts as a diffusional barrier for high energy particles due to the large
difference in atomic masses. 
The heavier atoms are predominantly ejected to the overlayer at the beginning of 
the thermal spike while the light atoms are injected to the bulk at the beginning of the
cooling period (in Ti/Pt) or during the thermal spike with some time delay (Al/Pt).
\\

61.80.Az 	Theory and models of radiation effects\\
61.80.Jh 	Ion radiation effects \\
61.82.-d 	Radiation effects on specific materials\\
61.82.Bg 	Metals and alloys\\
66.30.-h 	Diffusion in solids (for surface and interface diffusion, see 68.35.Fx)\\
79.20.Rf 	Atomic, molecular, and ion beam impact and interactions with surfaces\\
81.40.Wx 	Radiation treatment (particle and electromagnetic) (see also 61.80.-x Physical radiation effects, radiation damage)\\

\end{abstract}
}

\section{Introduction}

 The rapid growth in the use of ion implantation and ion beam processing of materials
in the semiconductor industry and elsewhere, such as ion-beam sputtering,
cleaning, smoothing, depth profiling, etc. \cite{Gnaser}, or
global response of solids under irradiation (e.g. phase instability in driven alloys) \cite{Bellon,Okamoto},
 makes it important to understand the radiation effects in materials and the displacement mechanism
of atoms during particle irradiation.
Despite the tremendous knowledge available in this field it is still not well understood
how to manipulate thin films, nanocrystalline materials, etc. in a proper way under ion irradiation driven conditions \cite{Samwer,Enrique}.  

  Ion-beam assisted processing of materials always leads to ion-beam mixing (IM), hence it is a topic
of a considerable technological interest. One of the greatest challenges is to keep
IM in a controllable fashion and to drive systems to a desired atomic configuration.
IM is oftenly used for amorphization, when ion bombardment drives the system far
from its equilibrium \cite{Okamoto,Samwer}.
Ion beam bombardment in mechanically stressed thin films and radiation-induced viscous flow are also hot topics of the area
\cite{Mayr}.

 Two types of phenomenological models have been developed to describe 
ion beam mixing.
The ballistic model assumes only the kinematic properties of the materials and the elastic collisions of energetic
particles (recoils) hence the deposited energy depends on the relative masses
of the colliding atoms \cite{Sigmund}. The effectivity of ballistic mixing (BM)
will depend on then the concentration of recoils with relatively
short mean free path which we call hot atoms.
BM tends to randomize atomic configurations leading to disordering of crystalline ordered
phases even in precipitating systems \cite{Enrique,Roussel}.
The slowing down of energetic particles, however,
leads to energy deposition and to local melting.
When a local volume is heated up induced by the slowed down
incoming ion and by the recoils, a thermal spike is created \cite{AverbackRubia}.

  The ballistic random  diffusion process \cite{Roussel} might also be active still during the 
TS and above a threshold energy leads to the formation of stable Frenkel pairs.
This results in
a supersaturation of the point defects and the enhancement of thermal diffusion \cite{Roussel}.
The competition between thermally activated and ballistic diffusional mixing might be a key
ingredient of IM \cite{Bellon,Roussel,Martin}.
In Martin's ballistic diffusional theory
atomic interdiffusion is influenced by atomic collisions and thermodynamic forces \cite{Martin}.
The latter is associated with thermally activated jumps of point defects.

 In addition,
thermochemical forces could be set in and influence thermally activated
atomic transitions \cite{AverbackRubia}.
In ref. \cite{Gades} it was explicitly demonstrated by MD simulations that within an
artifical Cu/Cu bilayer system the heat of mixing has a considerable effect on broadening.
Therefore the coexistence of chemical interdiffusion and BM could easily be
occurred (chemically guided collisional mixing \cite{Kelly}).

  The influence of thermodynamic driving forces on IM is summarized by Cheng \cite{Cheng} within a series of bilayer samples. 
Although they found a considerable scatter in the data the trend seems to be unmistakable: the amount of IM depends linearly on $\Delta H_m$ 
as it was proposed by Johnson and co-workers \cite{Johnson}, and the slope of
the mixing versus heat of mixing ($\Delta H_m$) curve is proportional to the cascade temperature ($\sim
10^4$ K) \cite{AverbackRubia}.
This TS model is based on the Vineyard model developed for thermally activated 
atomic transitions \cite{Vineyard}, and relates IM to the thermodynamic quantities,
the cohesive energy ($E_{cohes}$) and $\Delta H_m$.
The model sharply distinguish between ballistic and TS mixing.
The chemical interdiffusion model is supported in a number of ways experimentally \cite{AverbackRubia,mix_exp}, however, mostly at high ion energies (several hundreds keV or even more).
The TS concept assumes that following a short cascade period a stable liquid phase occurs
which persists up to several ps and no high energy particles (recoils) survive the cascade period.

 The mechanism of IM, is, however, less extensively studied at lower energies experimentaly
\cite{Gnaser} and by atomistic simulations \cite{AverbackRubia,TS,Colla,Gades,Nordlund,Sule1,Sule2}.
What is more or less
accepted now that the thermal spike plays an essential role in IM
above $1$ keV ion energy \cite{Gnaser}. Although the magnitude of TS IM is uncertain with the increasing
ion energy.
While the TS model assumes that all the atomic transport occur during the TS, other
authors suggest that IM takes place afterwards, in the relaxation period, because of the motion
of radiation-induced defects \cite{Kelly}.
 In our recent communication we studied the effect of $\Delta H_m$ on IM in Ti/Pt by
MD simulations and found no apparent effect of chemical forces on intermixing 
\cite{Sule1}.
This is in contrast with other low-energy MD studies where the effect of $\Delta H_m$ was apparent
\cite{Gades,Nordlund}.
We discuss in detail in the Discussion section the possible reason of the discrepancy.

 Low energy ion mixing (several keV or less) is even more complicated because in this case
the ion-solid interaction takes place close to the free surface and the
structural rearrangements at the surface and the bulk are coupled \cite{Sule1,Sule2}.
The atomic relocation processes are affected by the proximity of the surface
leading to a cooperative atomic transport as well as to surface cavity growth \cite{Sule2}.

  In this paper we report on molecular dynamics simulations up to 7 keV ion energy in two
metal bilayers. As in ref. \cite{Sule1} we found no apparent dependence of IM on $\Delta H_m$, although we explicitly demonstrate that an extended liquid ensemble (TS) is present at the interface. Contrary
to this a strong influence of the atomic mass ratio on interfacial mixing is explored.
Hence the role of the TS model can be questionized below $7$ keV ion energy.
We have chosen two system for detailed studies of IM: the already at low energies well studied
Ti/Pt \cite{Sule1} and the Al/Pt systems. These bilayers are chosen because of the considerable
difference in the atomic masses of the constituents and as such are suitable for the
study of the effect of the mass ratio on IM \cite{Sule2}.  
We examine the relative role of ballistic mixing vs. thermally activated vacancy mechanism
as it was given in the macroscopic continuum model of Martin \cite{Bellon,Martin}.
A different picture could be emerged if 
both processes are understood as fast diffusion processes in an atomistic description
which act against each other \cite{Bellon,Enrique}.

 The structure of this paper is as follows. First we outline the simulation method
used and define certain expressions used frequently throughout the paper.
In the discussion section we focus on the problem of the coexistence of the ballistic mechanism
and the thermal spike.
We also demonstrate the decoupling of the ballistic and thermally activated diffusional
mixing processes.
We discuss and explain the presence of a backscattering mass effect of the light particles
at the interface.


\section{The setup of the simulation}

Classical molecular dynamics simulations \cite{Allen} were used to simulate the ion-solid interaction
using the PARCAS code developed by Nordlund {\em et al.} \cite{Nordlund_ref}.
Here we only shortly summarize the most important aspects.
The variable timestep 
and the Berendsen temperature control is used \cite{Allen}. The bottom layers 
are held fixed and used as heat sink (heat bath) to maintain the thermal equilibrium of the entire
system.
The detailed description of other technical aspects of the MD simulations are given in \cite{Nordlund_ref} and details specific to the current system in recent
communications \cite{Sule1,Sule2}.

  The Ti/Pt sample consists of $328000$ atoms for the interface (IF) system
with $16$ Ti top layers and a bulk which is Pt.
The lattice constants for Pt is $a \approx 3.92$ \hbox{\AA} and for Ti $a \approx
2.95$ and $c \approx 4.68$ \hbox{\AA}.
At the interface (111) of the fcc crystal is parallel to (0001) of the hcp
crystal
and close packed directions are parallel \cite{Sutton}.
The interfacial system as a heterophase bicrystal and a composite object of 
two different crystals with different
symmetry are created as follows:
the hcp Ti is put by hand on the (111) Pt bulk and various structures are probed
and are put together randomly. Finally that one is selected which has the smallest
misfit strain prior to the relaxation run.
The remaining misfit is properly minimized below $\sim 6 \%$ during the relaxation
process so that the Ti and Pt layers keep their original crystal structure and we get an
atomically sharp interface.
The corresponding Ti-Ti and Pt-Pt interatomic (first neighbour) distances are $2.89$ and $2.73$
\hbox{\AA} at
the interface.
The Ti (hcp) and Pt (fcc) layers at the interface initially are separated by $2.8$ {\AA} and allowed freely to relax during the simulations. The variation of the equilibrium Ti-Pt distance within a reasonable $2.6-3.0$ \hbox{\AA} interval in the interatomic potential does not affect the final results significantly. We find the average value of $d \approx 2.65$ \hbox{\AA} Ti-Pt distance in the various irradiation steps and also in the nonirradiated system after a careful relaxation process. We believe that the system is properly relaxed and equilibrated before the irradiation steps.

  Another sample (Al/Pt) consists of 280000 atoms ($200 \times 200 \times 100 \hbox{\AA}$ for the interface (IF) system
with 6 fcc-Al top layers and a bulk which is fcc-Pt both with (111) orientation.
The interface is (111) oriented.
We find the size of the system is large enough to limit the influence of
the boundaries on the internal material transport.
The embedded atomic potential of Cleri and Rosato \cite{CR} is used and modified
to reproduce the interatomic distance for Al-Pt found in the AlPt alloy.
The heat of mixing for Al-Pt is also fitted to the experimental values \cite{Miedema},
although we find no significant dependence on it in accordance with earlier findings
in the TiPt system \cite{Sule1}. 
This system size is capable of accepting as large as 8 keV ion bombardment.
In this paper we show results at 6 keV ion energy.
The entire interfacial system is equilibrated prior to the irradiation
simulations and the temperature scaled softly down towards zero at the
outermost three atomic layers during the cascade events \cite{Nordlund_ref}.
We believe that the system is properly relaxed and equilibrated before the irradiation steps.
We found no vacancies at the interface or elsewhere
in the system after the relaxation procedure.
  We visually checked and found no apparent screw dislocations, misorientations or any
kind of undesired distortions (in general stress-generator dislocations) at the interface or elsewhere in the system which could result in
stress induced rearrangements in the crystal during the simulations.

 Collsion cascades were initiated by giving an Ar atom a kinetic energy of 1-7 keV at
7 deg impact angle (grazing angle of incidence) and at randomly chosen impact positions in the
central region of the free surface.
The SRIM96 electronic stopping power \cite{SRIM} was used to describe energy loss to electrons for all
atoms with a kinetic energy higher then 10 eV.

   We used a many body potential, the type of an embedded-atom-method given by Cleri and
Rosato \cite{CR}, to describe interatomic interactions.
This type of a potential gives a very good description of lattice vacancies, including migration
properties and a reasonable description of solid surfaces and melting \cite{CR}.
Since the present work is mostly associated with the elastic properties,
melting behaviors, surface, interface and migration energies, we believe the model used should be suitable for this study.
 The Ti-Pt and Al-Pt interatomic potential of the Cleri-Rosato \cite{CR} type is fitted to the experimental heat of
mixing of the corresponding alloy system. Further details are also given elsewhere \cite{Sule1,Sule2}.
  To obtain a statistics, several events are generated ($\sim 5-10$), and the typical events are
analyzed in the paper. Therefore no averaging is carried out over the events.
The variation of mixing as a function of the impact position and various events is shown
in refs. \cite{Sule1,Sule2}.

 Following the simulations we analyse the generated structural data base (movie file) which contains
the histories of the atomic positions using a code written for this purpose.
The analysis includes the identification of vacancies, liquid and mixed atoms, mean free path
of recoils and the extraction of information on local temperature of high energy particles
using the results of liquid analysis.

  Those atoms are recognised mixed (intermixed) which moved at least 0.5 monolayers ($\sim 1.4$ \hbox{\AA})
across the interface.

  The recognition of {\em vacancies} is done using a Voronoy polyhedron analysis
together with a Wigner-Seitz occupation analysis \cite{Nordlund_ref,Nordlund_PRB97}.
Within this approach we interpret voids, cavities or craters as vacancy clusters \cite{Sule2}.
Another problem is that in liquids, such as appears during the thermal spike,
the definition of vacancies is rather plausible since the empty cells within a liquid
are highly mobile. Hence during the TS we monitor only formally the time evolution of
the number of vacancies. This is important to indentify the appearance of the
supersaturation of vacancies, which is in fact the monitoring of the change of
the concentration of the empty cells within a liquid zone.

 {\em Liquid analysis} is also carried out and we call an energetic atom liquid if its
and its first neighbours kinetic energy exceed $k T_m$ where $T_m$ is the melting temperature.
Important requirement for a liquid atom is that
it should be a member of a real liquid ensemble, hence an individual fast moving hot atom or a recoil
can not be considered as liquid.
Though the quench rate of the liquid phase is large, it can persist normally up to  
several ps. In certain cases (e.g. Al/Pt at 6 keV) we observe an extraordinary long
lifespan for the local melt reaching $20$ ps or even more.
It turned out that one has to be careful when defining the liquid atoms.
Strictly speaking we denote those atoms liquid for which $T_m < T_{local} < T_{TS}$, where
$T_{local}$ and  $T_{TS}$ are the local temperature
and an arbitrary upper limit for the TS temperature, respectively.
The time dependent local temperature of the $i$th high energy particle ($T_{i,local}(t)$) can be given as follows:
\be
\frac{1}{2} m_i v_i^2(t) = \frac{3}{2} k T_{i,local}(t),
\ee
where $v_i$ and $m_i$ are the atomic velocity and mass of the $i$th high energy particle and $k$ is the Boltzmann constant.
We chose $T_{TS} \approx T_m+1000$ K, which seems to be a reasonable choice, because an atom
with $T_{local} >> T_{TS}$ is moving through the liquid/solid interface and migrates rapidly
in the interstitial space (recoil).
We also carry out recoil and hot atom analysis. For a hot atom we use the definition
$T_m+1000 < T_{local}$. 
We find a considerable amount of hot atoms both in Ti/Pt and in Al/Pt at various ion energies. In Al/Pt under 6 keV bombardment the high energy particles persist up to $20$ 
ps.
We use the notation recoil for those individual high energy particles, which can have extraordinarily
high local temperature and very short lifetime ($\tau_{recoil} < 3-500$ Fs) with long mean free path
($> 10$ \hbox{\AA}).
Hot atoms, however, can be taken as highly mobile high energy particles with medium mean free path
(around several times the lattice constant).
Therefore the hot atoms might thermalize a given region with a size of which is comparable with the 
mean free path of the hot species.
The kinetic energy of a hot atom is below the threshold of a stable Frenkel pair formation
energy therefore the ballistic jumps of hot atoms lead to subthreshold events which, however, 
are shown to contribute noticeably to the evolution of the damaged structure during IM (e.g. close Frenkel-pair recombinations, etc.) \cite{Wollenberger}.
Another basic difference between a recoil and a hot atom is that the former is
moving "freely" in the solid while the latter is a weakly bounded particle,
hence it interacts weakly with its neighbourhood.
It should also be mentioned that a brief definiton of hot atoms has given in
ref. \cite{Nordlund_ref} where those energetic atom are considered hot which
have experienced less then 3-5 lattice vibrations.
In this paper we prefer to select the energetic particles on the basis of their
local temperature regime.

  In order to avoid confusion we define the expressions diffusion and migration.
These phenomena are widely used in the literature for different processes \cite{Shewmon}, although
their meaning often is not clearly defined. Even more difficult to define the
precise meaning of radiation enhanced diffusion (RED).
In the present article in general we use the expression diffusion for any kind of
atomic migration processes which are taken through a solid.
Usually diffusion is used for slow atomic migration process on a ms or longer timescale.
Therefore we use the "fast" or "ultrafast" diffusion expressions if the timescale
is shorter \cite{Okamoto,Martin,Sutton}.
If we do not specify the duration of the process, we prefer to use atomic migration.
Also we use the notion RED for those atomic transport processes which are taken place
within tens or hundreds of ps. This is typically the timescale of MD simulations. 
Sometimes the RED is used for any kind of atomic transport phenomena in the literature
which appears under irradiation.
We prefer not to use RED for diffusion on a longer timescale (ms or longer) which is in our opinion can not be
distinguished from diffusion under parent conditions \cite{Shewmon}.
The expression ballistic diffusion \cite{Roussel} can also be used instead of
(ultra)fast diffusion.


\begin{figure}[hbtp]
\begin{center}
\includegraphics*[height=5cm,width=6cm,angle=0.0]{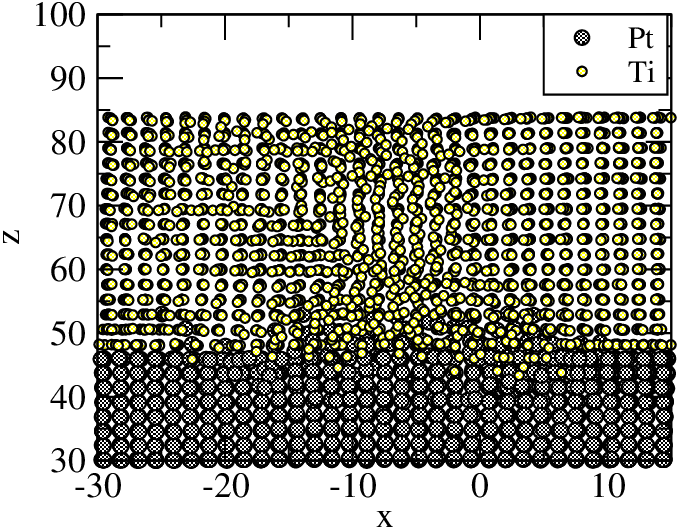}
\includegraphics*[height=5cm,width=6cm,angle=0.0]{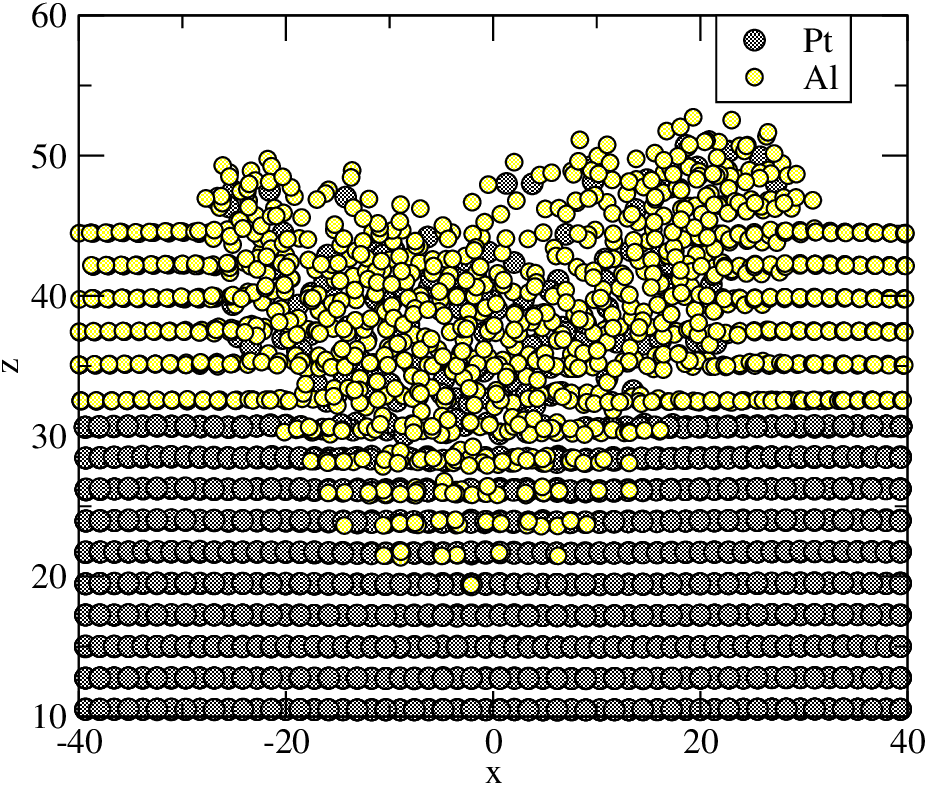}
\includegraphics*[height=5cm,width=6cm,angle=0.0]{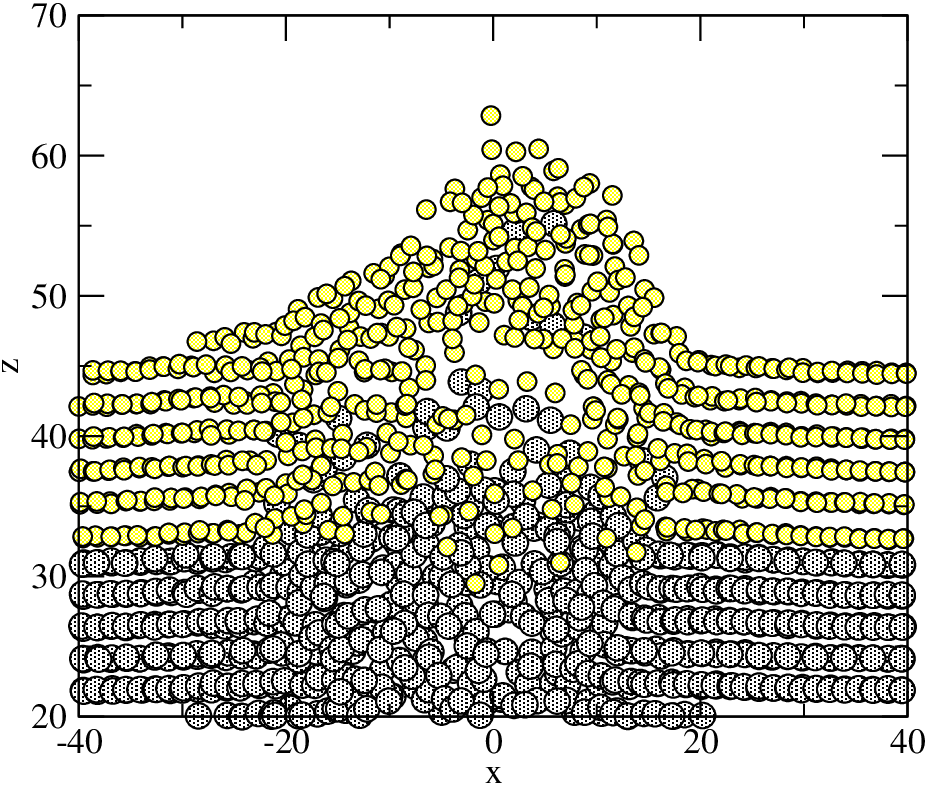}
\caption[]{
The cross sectional view of the Ti/Pt system after a 7 keV irradiation
at the end of the relaxation process.
The overlayer is Ti and the substrate is the Pt.
{\em Middle panel}: The cross sectional view of Al/Pt after 6 keV
irradiation and 55 ps after the ion impact.
The thickness of the cross-sectional slabs are $10$ \hbox{\AA} in both cases.
{\em Lower panel}: The cross sectional view of Al/Pt after 6 keV
at 2.5 ps.
}
\label{xz}
\end{center}
\end{figure}


\section{Results and discussion}

 First we focus on the lack of the dependence of IM on chemical forces and then
we try to address a ballistic scenario for mixing without the usage of the TS concept
\cite{Cheng}.
 Already we have presented in our previous communication \cite{Sule1}
that we find no dependence on $\Delta H_m$ in Ti/Pt, although we found
extensive IM at 1 keV ion energy.
The Cleri-Rosato many body potential can be fitted to the experimental heat of mixing
or can be varied in order to account for any values of $\Delta H_m$.
Hereby we do not give the technical details of the variation of $\Delta H_m$ in the 
interatomic potential and refer to another ref. \cite{Sule1}.
Furthermore, we explored the decoupling of the intermixing of the constituents:
first the bulk Pt atoms are ejected to the top layers and at the end of the TS
the top layer Ti atoms are injected to the Pt bulk.
We had no clear cut explanation for this unexpected behavior of this system.
In this paper we would like to demonstrate that basically we find the same effects
at higher energies.

 In the next subsections we summarize the results obtained and outline certain possible
scenarios which might help to understand IM.
These are the following: competing dynamics between ballistic diffusion and 
vacancy saturation induced reordering \cite{Martin}.
We explain the asymmetric mixing by a backscattering effect of the light atoms
at the interface which delays their intermixing. 
We try to compare our results with experiments obtained for glassy and amorphised metals
in order to point out the parallels.
In any of the outlined mechanisms we discuss the essential role of hot atoms in IM.

\subsection{Intermixing and liquid analysis}

\begin{figure}[hbtp]
\begin{center}
\includegraphics*[height=4.5cm,width=6.cm,angle=0.0]{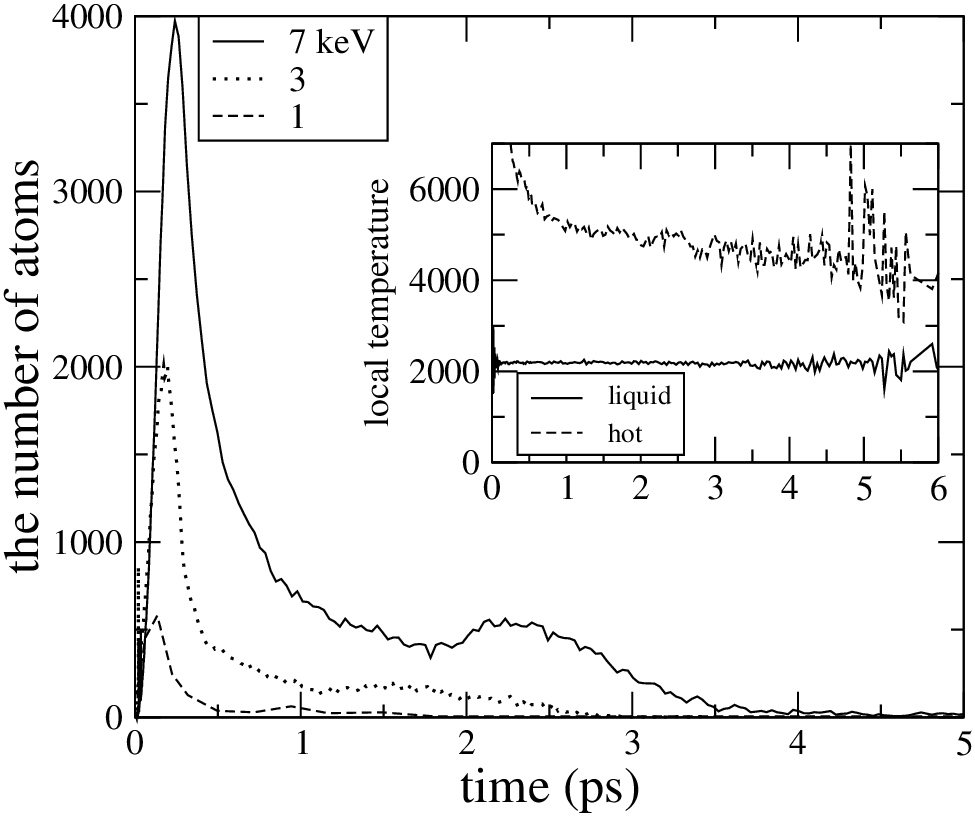}
\includegraphics*[height=4.5cm,width=6.cm,angle=0.0]{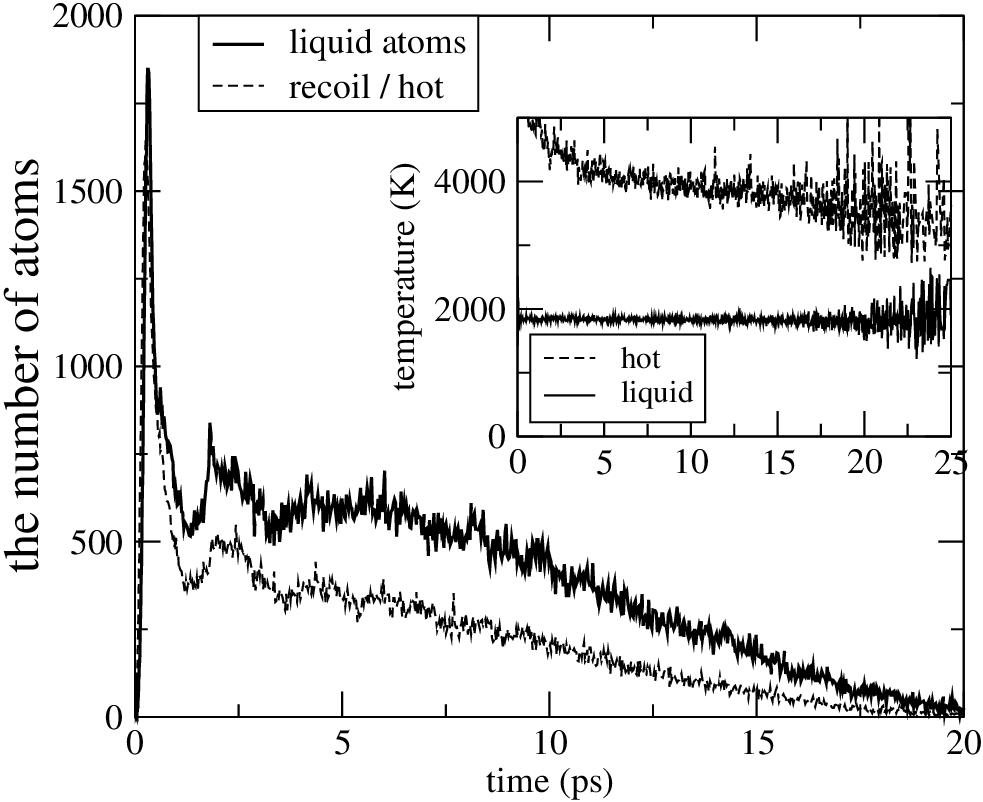}
\caption[]{
{\em Upper panel}:
The number of liquid atoms in Ti/Pt
 at various ion energies up
to 7 keV and the local average temperature (K) at 7 keV ion energy (inset).
{\em Lower panel}:
The number of liquid, hot atoms and recoils in Al/Pt as a function of the time (Fs).
{\em inset:} the time evolution of the local temperature in the liquid phase (K),
the number of the liquid atoms, recoils and hot particles.
}
\label{liquid}
\end{center}
\end{figure}

  The cross-sectional view of the irradiated Ti/Pt sample at 7 keV is shown after 30 ps
of the ion impact in FIG ~(\ref{xz}) which clearly demonstrate that the interface is
damaged and large amount of intermixing occured and remained there built-in.
A complete animation is available on the world wide web \cite{www}.
Liquid analysis provides us the information that the number of liquid atoms
is increased heavily with the increasing ion energy (FIG ~(\ref{liquid})).
Therefore there should be present a liquid phase which behaves as a real TS.
Contrary to this we find no apparent dependence on $\Delta H_m$. 
Even if $\Delta H_m \ge 0$, considerable mixing occurs instead of segregation.
This is in direct conflict with the results obtained in ref. \cite{Gades}.
We explain the discrepancy with the following.
Gades and Urbassek studied the effect of $\Delta H_m$ on IM in an artifical system
of Cu/Cu bilayer \cite{Gades}. They varied $\Delta H_m$ within this system in which the first six
layers consist of natural Cu and the substrate of a modified Cu which has a positive
or negative heat of mixing towards Cu.
They found a strong effect of $\Delta H_m$ on broadening.
We attribute the presence of a chemically guided mixing to the 1:1 mass ratio, to the lack of lattice missmatch and and to the same cohesive energy in the
bulk and in the overlayer.
Within their computer experiment the only variable was $\Delta H_m$, while in a real 
system this is far not the case. 
In a general bilayer system couple of other parameters set in which may suppress the effect
of $\Delta H_m$.
The relocation energy of an atom in a solid is affected
by a couple of other parameters \cite{Allnat}.
E.g. in the case of Ti/Pt and Al/Pt there is
a considerable mass ratio, difference in cohesive energy and in the melting points of
the constituents not to mention the lattice missmatch.
Another problem with chemically guided IM is that $\Delta H_m$ could only affect
atomic transport processes when the atoms are moving relatively slowly with ordinary
diffusion. Fast moving particles are only weakly influenced by chemical forces and the chemical
environment affects only those species which are bounded at a lattice site.
We will show later on that the concentration of ballistic particles (hot atoms)
is sufficiently high in Ti/Pt and in Al/Pt during the TS to supress the effect of $\Delta H_m$.
Although the chemical interaction between liquid atoms is stronger, however, their
mobility might still be sufficiently high to be affected only weakly by $\Delta H_m$.
This is a subject, however, what should be carefully examined in the future and which goes
beyond the scope of the present article.

\begin{figure}[hbtp]
\begin{center}
\includegraphics*[height=4.5cm,width=6.cm,angle=0.0]{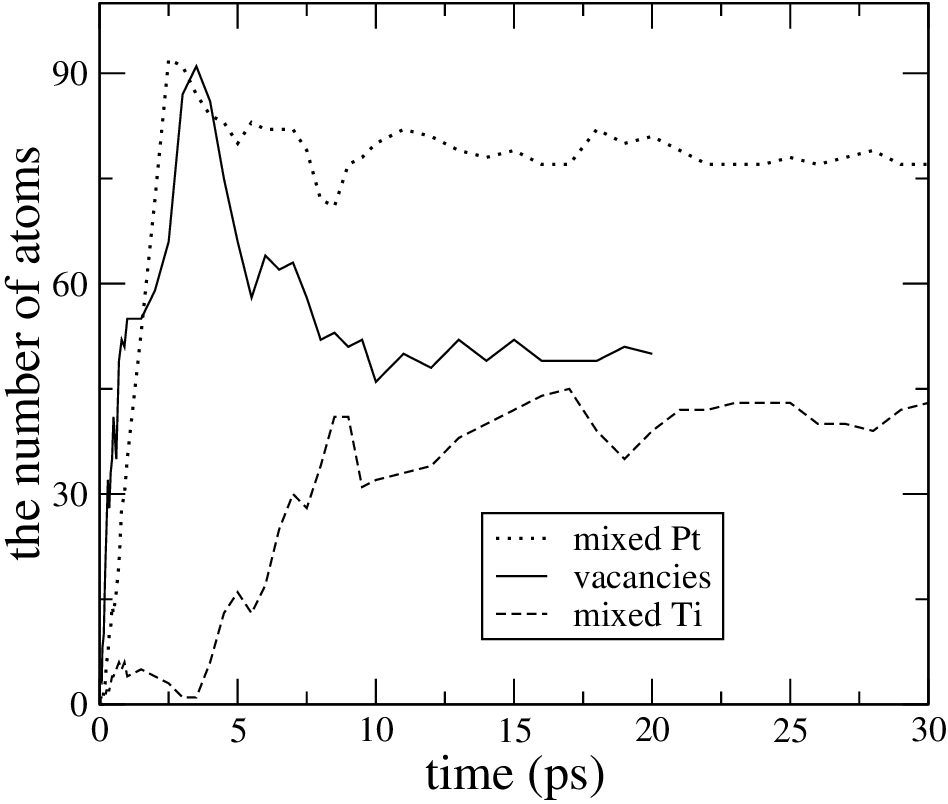}
\includegraphics*[height=4.5cm,width=6.cm,angle=0.0]{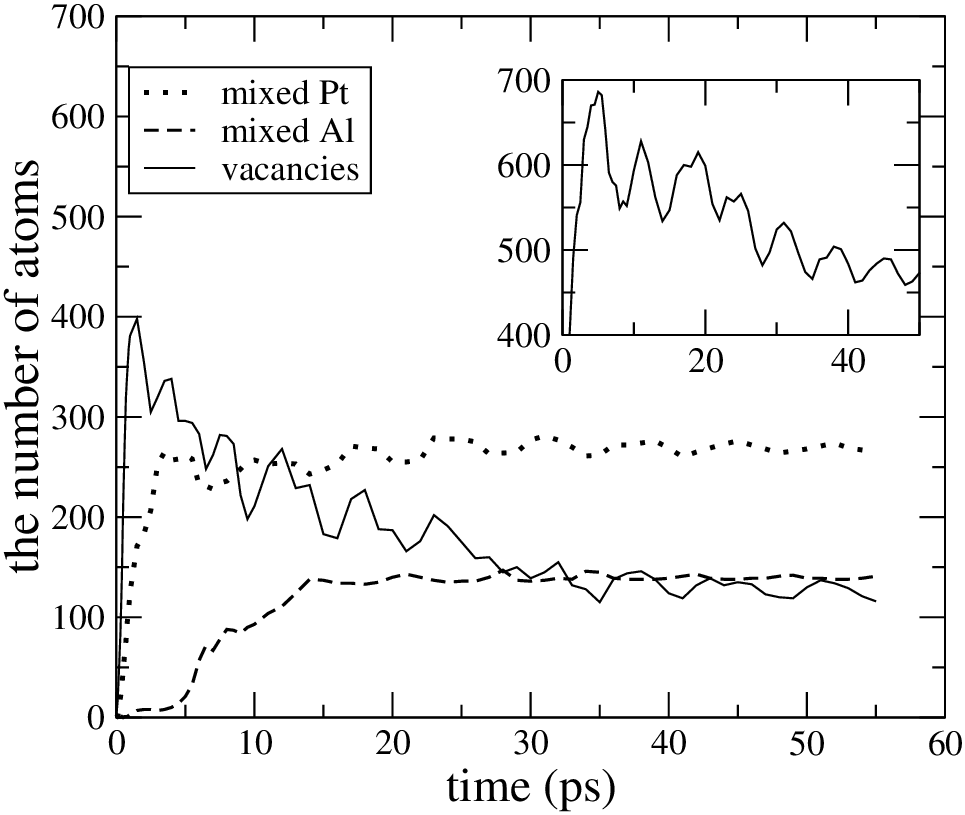}
\caption[]{
The time evolution of mixing and vacancy formation in Ti/Pt (upper figure) at 7 keV ion energy. Mixing profile in Al
/Pt (lower figure) together with the time evolution of vacancies at 6 keV ion energy.
The number of adatoms are also shown in the inset of the lower figure as a function of the time
 (ps).
}
\label{mixing}
\end{center}
\end{figure}

  We find a strong asymmetric behavior in mixing (FIG ~(\ref{mixing})).
The decoupling of the mixing of the top layer constituent from the TS period
is also apparent (FIG ~(\ref{mixing})).
Asymmetric mixing behavior has also been observed in Ni/Ag \cite{Colla} and in Cu/Ni or in Cu/Co
\cite{Nordlund}.
In our case first the heavier element Pt mixes and the intermixing of the light constituent Al or Ti 
is initiated at 5 ps. 
The mixing of Ti (Ti/Pt) continues after the TS period hence
no liquid atom is required for the mixing of Ti.
Pt mixing peaks at around 4 ps.
We find no intermixing hot or liquid Ti atoms. 


The number of vacancies $N_{vac}$ exhibits a sharp peak at 4 ps (upper FIG ~(\ref{mixing})) which 
corresponds to a highly diffuse liquid state in the Ti overlayers and
approximately the number of vacancies is 3 \% of the number of the liquid atoms
which indeed leads to the sudden decrease of the atomic density (not shown).
The intermixing of the Ti atoms might be initiated by the saturation of Pt vacancy concentration at around 4 ps.
The supersaturation of point defects results in enhanced atomic mobility \cite{Bellon}:
a ballistic jumps are increasingly probable in the direction of nearby vacancies.
Replacement collision sequences might be inititated in this way (via subtreshold collisions)
which enhances atomic mixing \cite{Bellon}.

  The results of the liquid analysis are also shown for Ti/Pt and Al/Pt at 7 and 6 keV
in FIG ~(\ref{liquid}). 
The TS persists up to $\sim 20$ ps in Al/Pt, which is an unusually long lifetime. 
The longest reported lifetime for a TS ($\tau_{TS}$) is 14 ps for Au at a similar ion energy regime in a 5 keV cascade event \cite{Nordlund_PRB97}.
For pure Al and for Pt a much shorter $\tau_{TS} \approx 4$ ps is obtained \cite{Nordlund_PRB97}. 
Also, in pure elements one can find no damage at the end of the simulation hence the restoring forces (recrystallization) are strong \cite{Sule2}.
As already mentioned above,
in the bilayer systems, however, the damage rate is high, the recovery process is uncomplete within the timescale of the MD simulations.
Moreover the accumulated damage rate is so high that it might also be remained built-in
on a longer timescale.

  In case of Al/Pt the average local temperature far exceeds the melting point of Al ($\sim T_m \approx 982$ K) value and is stabilized around
$\sim T \approx 1800$ K.
When the number of liquid atoms is less then a critical
value ($\sim 50$), the local temperature starts to oscillate due to the 
abrupt condensation of the liquid atoms at the end of the TS.
The TS exhibits in general an
ultrarapid quenching rate ($\sim 10^{14}-10^{15}$ K/s, $\sim 100-1000$ K/ps) which is much larger then
the values are given in the literature $\sim 10^{12}$ K/s for supercooled metallic glasses \cite{Samwer}),
leading to a quenched metastable solid solution at the interface with compositions lying beyond the 
equilibrium solubility limit. 
Others also propose that fast quenching, occurring the estimated rate of about
$10^{14}$ K/s provides ideal conditions to nucleate metastable phases, including
the amorphous one \cite{Ossi} or to the clustering of vacancies \cite{Sule2}.
We find also a broad amorphous interphase as demonstrated in the middle FIG ~(\ref{xz}).

  Glassy metals exihibit a polymorphic melting point falls
below the ambient temperature and below the glass transition temperature of the
liquid phase.
Under these circumstances the oversaturated solution undergoes a crystal-to-glass
transformation \cite{Samwer}.
We find also a polymorphic melting point (liquid temperature) which lies between the parent melting point of
the constiuents.
The fast "cool" mixing of Ti might be induced by a glassy state saturated by vacancies.
The mixing of Ti is fast enough not to be uderstood by a simple vacancy mechanism \cite{Shewmon}.
Therefore we argue that a glassy sate might accelerate the diffusion of the Ti species.
The initialization of mixing of Ti coincides with the sudden fall of $N_{vac}$.
Hence the vacancy supersaturation initializes a fast intermixing which requires no
liquid atoms in Ti/Pt.
 In Al/Pt the situation is somewhat different. Here the vacancies tend to aggregate while
in Ti/Pt a diffuse vacancy saturation occurs with little or no vacancy clustering which
favors the formation of a glassy sate.

  The broad intermixed interface in Al/Pt 
(seen in FIG ~(\ref{xz})) is consistent with the ion-induced amorphization experiment carried
out on Al/Pt \cite{Gyulai,Buchanan}, in which a uniform mixing is achieved and confirmed the existence of an amorphous phase
at the interface over a large composition range.
In general it is well documented now that binary metal systems undergo solid-sate 
amorphization by interdiffusion or by other type of mixing \cite{Samwer}.

\begin{figure}[hbtp]
\begin{center}
\includegraphics*[height=4.5cm,width=6.cm,angle=0.0]{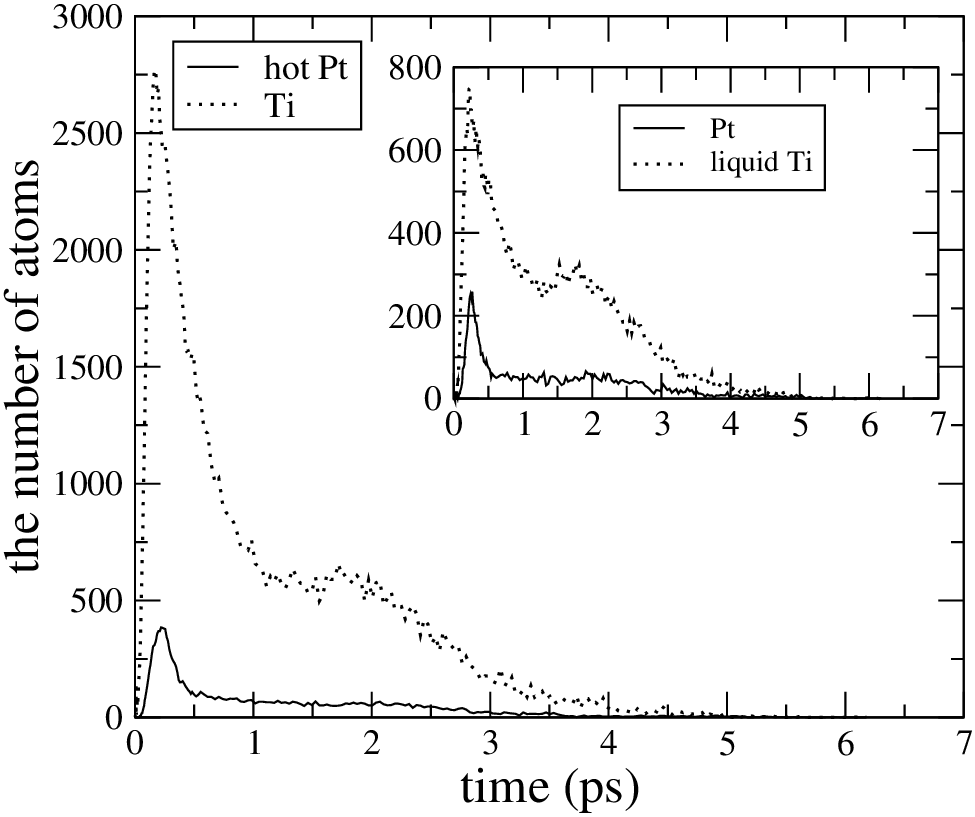}
\includegraphics*[height=4.5cm,width=6.cm,angle=0.0]{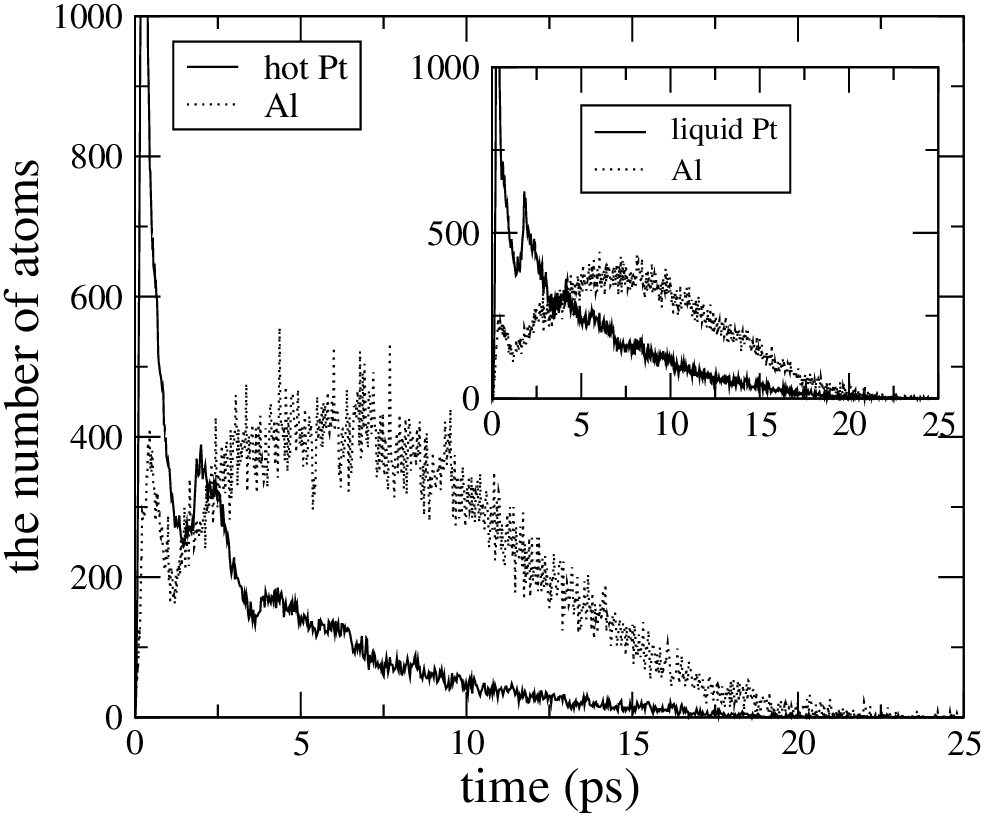}
\caption[]{
{\em Upper panel:}
The time evolution of the number of hot and liquid atoms (inset)
in Ti/Pt at 8 keV Ar$^+$ irradiation.
{\em Lower panel:}
The time evolution of the number of liquid (inset) and hot atoms in Al/Pt at 6 keV
ion irradiation.
}
\label{liqhot}
\end{center}
\end{figure}

 Another interesting feature of FIG ~(\ref{liquid}) is the time evolution of
recoils/hot atoms. The recoils dissapear at the end of the
cascade period ($\sim 0.3$ ps) \cite{AverbackRubia}. 
The sharp peak at $\sim 0.3$ ps corresponds to the recoils.
However, we find that a large amount of high energy
particles (hot atoms) survive the cascade cooling.
This is a rather surprising result and it shows that the ballistic period with hot atoms 
coexists with the TS in these bilayers. 
In the inset of lower FIG ~(\ref{liquid}) we show that the average temperature of
the hot atoms far exceeds $T_m$.
In FIG ~(\ref{liqhot}) we show the time evolution of hot and liquid
atoms separately for various atom types.
Not surprisingly the various atoms exhibit different time evolution
due to the large melting point difference.
The number of Pt hot and liquid atoms
delays rapidly while the number of Al energetic particles show a saturation at 5 ps.
It has been found that the solidification time of the cascade together with hot and liquid atoms
depends on the melting point of the material \cite{Nordlund,Nordlund_ref}.
Hence the intermixed Pt particles quench out more rapidly from the liquid zone then the Al atoms and remained mixed
atoms at the liquid zone boundary.  
Nevertheless it should be mentioned that in Ti/Pt the number of Pt liquid and hot 
atoms is considerably smaller then those of Ti because the displacement cascade
is localized in the overlayer and the Pt side of interface is hit only
slightly by recoils. This is because a relatively thick Ti overlayer is constructed
(with 16 monolayers).
 In Al/Pt the Pt atoms exhibit a second hot and liquid atom peak (the first one is due to the recoils)
at 2 ps which coincides with the end of the fast mixing of Pt atoms (FIG ~(\ref{mixing})).
Hence the hot atom peak is a mixing peak due to the ballistically mixing hot Pt particles.
In Ti/Pt we see only the broadening of the first recoil peak hence the increasing mass difference
might induce the development of a second recoil (hot) peak.

\subsection{Thermally activated diffusion and vacancies}

  In Ti/Pt we found that
the fast quench of the TS leads to extensive vacancy formation \cite{Sule1}, and
this is what induces the injection of Ti atoms to the bulk.
Hence we also support a vacancy mechanism for mixing with a light atom injection to the bulk.
The large concentration of vacancies is an indicative of a glassy interface \cite{Okamoto}.
It should be remarked that Pt intermixes to the top layers during the TS possibly
with a ballistic interdiffusion mechanism because we find a considerable amount of
mixing hot Pt atoms.
The interdiffusion of Pt leads to amorphous inter-layer growth at the original interface
and in this sense resemble to an anomalous fast diffusion \cite{Okamoto,Sutton}.
Moreover due to the elevated temperature of the TS, a thermally activated diffusion process might become
dominant.
Although at this moment it is difficult to determine whether thermally activated or ballistic
diffusion is the dominant atomic migration process during the TS.

  In Al/Pt we see that the number of vacancies peaks at 2.5 ps. A more elaborate
analysis of the structures at that time we see that vacancies are clustering 
and temporarily form a cavity (lower FIG ~(\ref{xz}).
Hence in Al/Pt the supersaturation of vacancies leads to clustering while in Ti/Pt
we have still a higly diffuse liquid state.
The most of the vacancies are located in the overlayer.
Comparing the heat of formation of a vacancy $\Delta H_V$
we find that $\Delta H_V \approx 1.31$ and $0.74$ eV for Pt and for Al 
\cite{Swalin}.
This is a marked difference and explains the large discrepancy in the number of vacancies
in these metals.
The value for Al is in the energy regime of the hot atoms hence the formation
of vacant sites throughtout the TS can be expected.
In Pt, however, we see that the vacancies dissappear rapidly (not shown).

  A competition between BM and vacancy saturation induced atomic transport (vacancy mixing)
might also be active in irradiated samples \cite{Roussel}.
The BM tends to randomize (disordering) the system leading to thermalization and vacancy
saturation. The thermally activated vacancy recombination or net flux through the
interface induces reordering \cite{Roussel}. 
In our samples we see the dominant feature of ballistic mixing in both cases for the
Pt atoms.
In Ti/Pt, however, the Ti atoms intermix with a vacancy mechanism afterwards the TS.
In Al/Pt both species migrate with a ballistic diffusion.

\begin{figure}[hbtp]
\begin{center}
\includegraphics*[height=4.5cm,width=6.cm,angle=0.0]{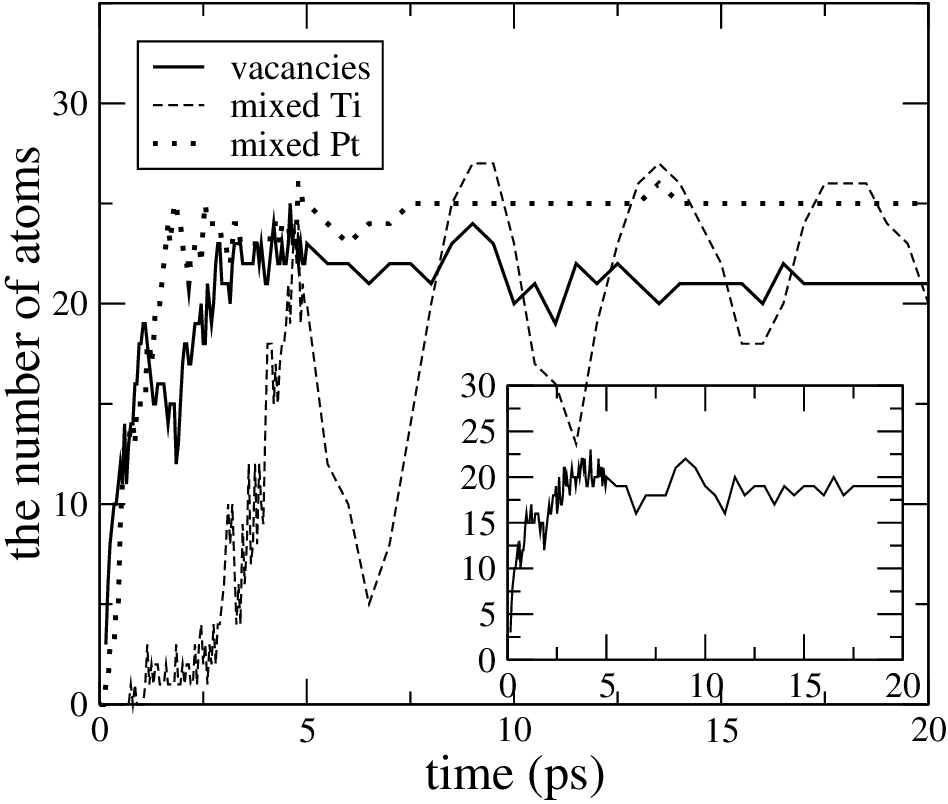}
\includegraphics*[height=4.5cm,width=6.cm,angle=0.0]{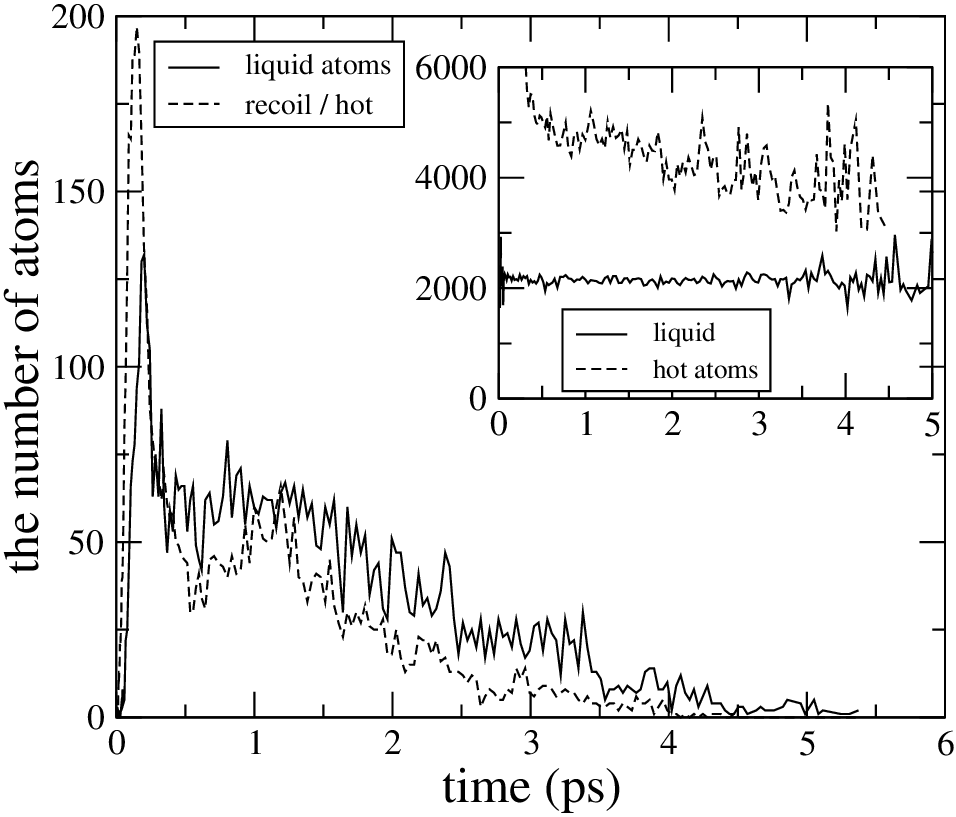}
\includegraphics*[height=4.5cm,width=6.cm,angle=0.0]{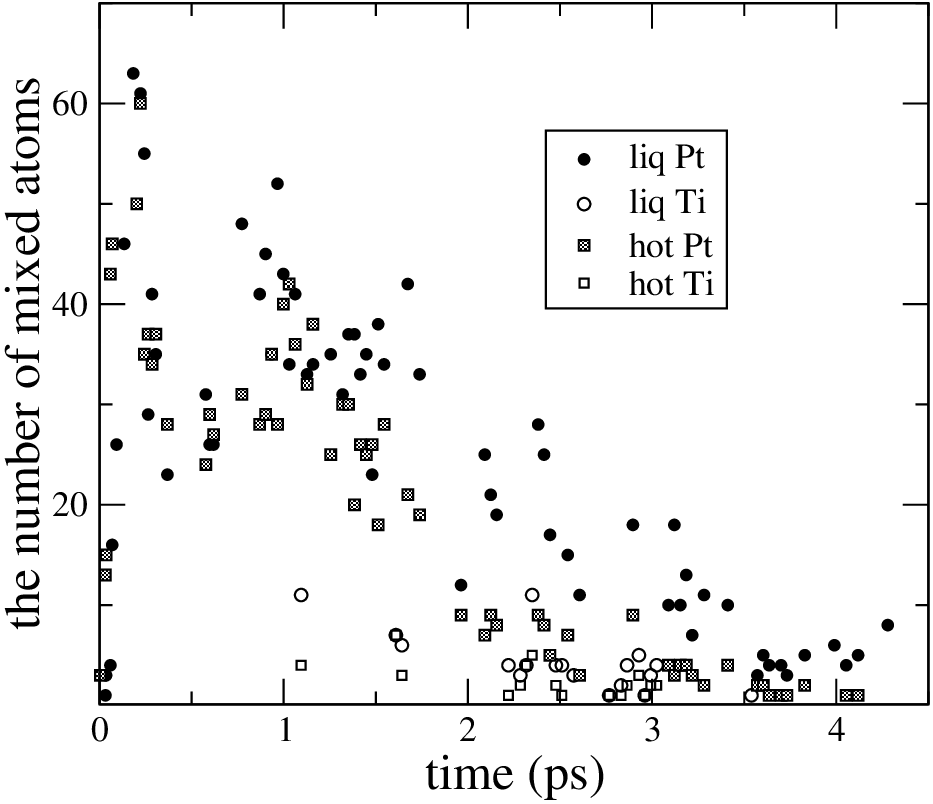}
\caption[]{
{\em Upper panel:}
The number of vacancies, adatoms and intermixed atoms
in Ti/Pt at 1 keV Ar$^+$ irradiation as a function of time.
The time evolution of the numbers of interstitials
is also shown in the inset.
{\em Middle panel:}
The time evolution of the number of liquid and hot atoms.
The average temperature (K) is shown of the liquid and hot atoms
as a funtcion of time (ps)
{\em Lower panel:}
The time evolution of
the number of mixed hot and liquid atoms.
}
\label{tipt_vacancy}
\end{center}
\end{figure}

\subsection{Mixing at 1 keV ion energy}

 In order to show that the long-lived hot atoms exist even at lower ion energies
we recall our results presented in ref. \cite{Sule1} and give further details
at 1 keV energy.
  We find the same effect for 1 keV bombardment of Ti/Pt, where the intermixing of Ti
coexists with the formation of vacancies \cite{Sule1} (upper FIG 
~\ref{tipt_vacancy}).
On a longer timescale in upper FIG ~(\ref{tipt_vacancy}) we see the oscillation of $N_{mix,Ti}$ (mixed Ti atoms)
which is due to a short distance movement of these atoms at the interface.
The vacancies in the interface tend to nucleate as it is demonstrated
in ref. \cite{Sule2} leading to a cavity growth under repeated irradiations.
Therefore we see a tendency to cavity formation in these systems which
is coupled to a mixing induced material transport \cite{Sule2}.
The number of interstitial atoms (inset upper FIG ~(\ref{tipt_vacancy})) exhibits
a similar time evolution to that of the vacancies which indicates us that the
mixing of Ti atoms is accomplished by vacancy exchange mechanism via interstitial
atoms. The vacancy mechanism is well established as the dominant mechanism of diffusion
in fcc metals and alloys and has been shown to be operative in many hcp metals \cite{Shewmon}.
However, it must be admitted that the mixing of Ti is quite fast. More then
20 Ti atoms intermix during 2.5 ps at the end of the TS and IM survives the
quench of the TS.

  Two types of intermixing mechanism can be seen which are more or less decoupled in time from each other:
First the fast ballistic diffuser specie
intermixes to the overlayer during the TS partly to the interstitial
positions or partly to vacancies left behind by overlayer atoms which migrated to the surface (adatoms).
\begin{figure}[hbtp]
\begin{center}
\includegraphics*[height=4.5cm,width=6.cm,angle=0.0]{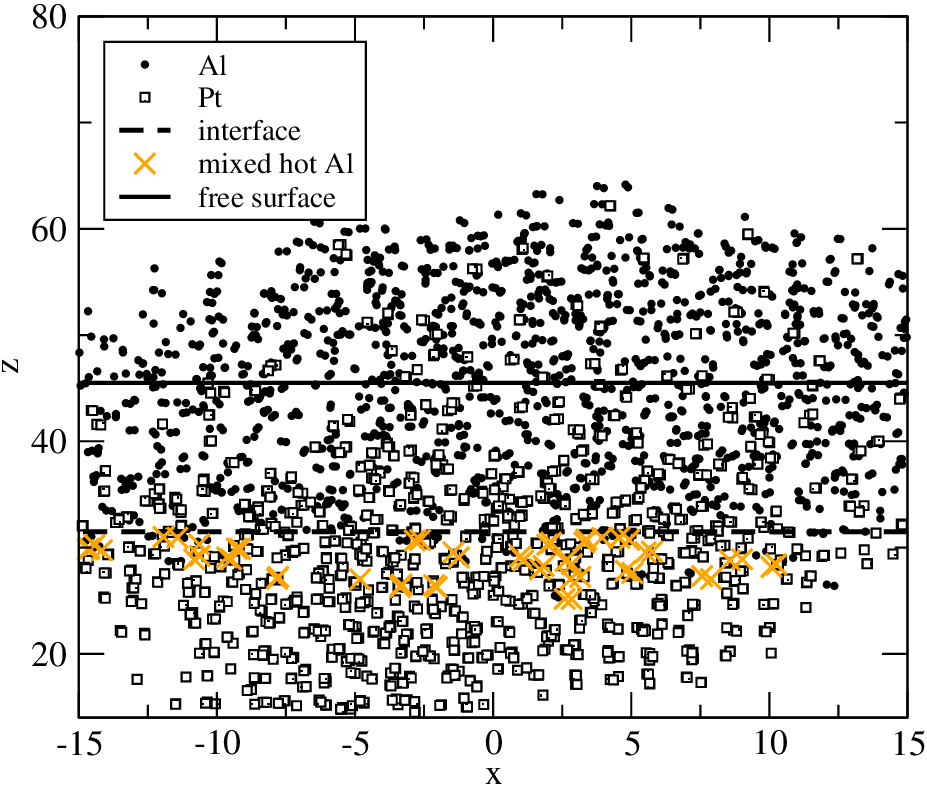}
\caption[]{
A crossectional slab of
hot atoms with a slab thickness of $10$ \hbox{\AA} at 5 ps in Al/Pt after 6 keV irradiation. Note that the Al recoils
are reflected at interface while Pt recoils enter the interface hence intermix
to the overlayers.
Intermixed hot Al atoms are also shown by light colored crosses after 5 ps.
}
\label{recoil_xz}
\end{center}
\end{figure}

After cooling of the TS adatoms return from the surface to the overlayer \cite{Sule1} and push Ti atoms
to the bulk. 
It should be remarked that in the case of Al/Pt the intermixing of Al starts still during
the TS, however, in both cases sets in at around 5 ps in the high ion energy cases (FIG 
~(\ref{mixing})).
The timescale of this fast atomic migration is very short (fractions of ps) in agreement with
the findings obtained phenomenologically \cite{Ossi}.

 In the lower FIG ~(\ref{tipt_vacancy}) the number of the intermixed
liquid and hot atoms are shown.
One can see that a considerable amount of energetic particles are mixed time to time
to the overlayers. Interestingly no or only few hot and liquid Ti atoms are mixed to the bulk, although
they are present during the TS in the overlayers. We explain this by the
backscattering of the energetic light atoms (Ti) at the interface due to the
large atomic mass difference (see later).
This phenomenon is already discussed briefly in ref. \cite{Gades}, although they
reached the conclusion that the effect of mass is small.
Probably that is the reason of the decoupling of the mixing of the constituents and the
delay of IM of the light atoms.
The mean free path of the hot atoms is around $\sim 10$ \hbox{\AA} and the
corresponding velocity of these mixing energetic particles is a robust $\sim 0.5$ \hbox{\AA}/Fs 
($\sim 50$ m/s).

\subsection{Backscattering at the interface and mass effect}

 The backscattering phenomenon is shown in detail in FIG ~(\ref{recoil_xz}) in Al/Pt
after 6 keV bombardment. It can clearly be seen that the Al atoms are reflected at the
interface until 5 ps, however, after 5 ps they start to move beyond the interface.
The reflection can be understood by simple kinematic reasons: the large mass difference
between Pt and Al results in the scattering of the light atoms.
Both species are reflected at the solid/liquid zone boundary.
It should also be mentioned that we find the same scattering phenomenon in Ti/Pt.
The treshold time for Al or Ti intermixing is still rather puzzling.
By the moment we attribute it to a vacancy supersaturation induced mixing which is strong
enough driving force for moving beyond the interface the light atoms at 5 ps.
The reflecting particles are confined within the thermalized hot zone leading to superheating of the
liquid region ($T_{aver} \approx 4000$ K).
Superheating effects under pressure in encapsulated metal clusters has also been reported
\cite{Banhart}.
The subject of superheating is, however, beyond the scope of the present article and hereby 
we simple note that the local melt of the TS is also under pressure put by the surrounding media hence superheating properties
might also be present.
The supersaturation of vacancies is also a consequence of the confinement induced thermalization.
The treshold time 5 ps for the light atom mixing coincides with the saturation of
the number of hot Al atoms (lower FIG ~(\ref{liquid})) and  with the sudden decrease in the number of vacancies (FIG ~(\ref{mixing})).
One might be assumed that at 5 ps the interface becomes sufficiently damaged
and saturated by vacancies to be permeable for light energetic atoms.
The temporal confinement of the cascade in the overlayer is also found in ref. \cite{Gades},
although they found much smaller mass effect since the smallest mass ratio was 0.5.
We expect the similar confinement effect in the reverse case when the bulk is formed from the
light atoms and the overlayer is from the heavy atoms.
In this particular case the light hot atoms are reflected from the interface from below, hence
the superheated region has no connection with the surface.

 In FIG ~(\ref{liqhot}) the number of Al hot atoms $N_{Al,hot}$ peaks at 5 ps which supports the
idea of interfacial barrier climbing at 5 ps: when $N_{Al,hot}$ peaks
Al hot atoms break through the interface and deposit their kinetic energy at the Pt side.
Therefore the interface is a ballistic barrier for interfacial mixing of light atoms even if they
are hot atoms. Therefore the barrier results in the phenomenon what we call {\em retarded ballistic mixing}
due to the semipermeable interface.
When the overlayers are sufficiently frustrated by vacancy supersaturation,
the light atoms break through the interface.
This process is much slower, however, then the fast ballistic diffusion of Pt.
The backscattering effect is also interesting in that point of view that
 the semipermeable interface divides the TS volume into to separated regions.
On the permeable side (Pt side) the liquid atoms are Pt atoms while on the Ti or Al side
the local melt is a mixture of the constituents.
Therefore there can be a concentration gradient towards the Pt side which can drive
the intermixing of Ti or Al according to the Fick's law \cite{Shewmon}.
In that case the driving force of ballistic mixing would be a concentration and
a thermal gradient which changes sign during the thermal spike (Al/Pt) or due to
the quench of the local melt (Ti/Pt).

\section{Conclusion}

 We demonstrated that a bilayer system exhibits enhanced intermixing properties and
an increased thermal spike lifetime when compared with the values obtained in pure elements.
We attribute this behaviour to the
sufficiently large difference in the atomic masses of the constituents (mass ratio).

 It is shown that, in a bilayer system under irradiation, ion beam mixing
is driven by a ballistic mechanism due to the unusually long lifetime of hot particles
in systems such as Al/Pt or Ti/Pt with a large mass ratio.
The interfacial mixing for the constituents are decoupled in time: predominantly
Pt mixes to the overlayer by ballistic diffusion (cascade mixing) and 
the light constituent (Al or Ti) starts to 
migrate to the bulk with some time delay.
We find that the energetic ligth atoms are backscattered from the interface
due to the large mass difference.
Hence the intermixing of them is delayed significantly.

  We find no sign of sensitivity of ion beam mixing to thermochemical properties (heat of mixing).
Instead strong random ballistic diffusional features rule the mechanism with a long lifetime of hot particles
hence with an enhanced cascade mixing properties.
A different mechanism is active for the light atom mixing in Ti/Pt and in Al/Pt. 
In the former bilayer Ti is injected to the Pt bulk afterwards the TS by a vacancy mechanism
while the Al atoms intermix during the TS via a time delayed (retarded) ballistic diffusion.
Therefore, there is a great difference in the details which should be precisely studied
case by case.
 Finally we would like to remark that in accordance with the theory of Martin \cite{Martin}
we find that ballistic diffusion and thermally activated vacancy movements 
seem to be suitable 
for understanding ion beam mixing.

\section{acknowledgment}
{\small
This work is supported by the OTKA grants F037710, T043704
and T30430
from the Hungarian Academy of Sciences.
Computer time at the Center of Scientific Computing (NIIF) in
Budapest is gratefully acknowledged.
}



\end{document}